# Nonadiabatic high-order harmonic generation from solids

GUANG-RUI JIA AND XUE-BIN BIAN*

*College of Physics and Materials Science, Henan Normal University, Xinxiang 453007, China, and State Key Laboratory of Magnetic Resonance and Atomic and Molecular Physics, Wuhan Institute of Physics and Mathematics, Chinese Academy of Sciences, Wuhan 430071, China*
*Corresponding author: xuebin.bian@wipm.ac.cn*

**We studied the multi-plateau high-order harmonic generation (HHG) from solids numerically. It is found that the HHG spectrum in the second or higher plateau is redshifted in short laser pulses due to the nonadiabatic effect. The corresponding FWHMs also increase, suggesting the step-by-step excitation process of higher conduction bands in the HHG process. Even order harmonics in the higher plateaus of HHG are present due to the break of symmetry in the $k$ space. It may stimulate related experimental work.**

http://dx.doi.org/xxxxx

High-order harmonic generation (HHG) from gases has been well studied. It has been used to generate attosecond laser pulses [1,2]. It is described by the three-step model [3]. The cutoff energy of HHG is around $I_p+3.17U_p$ ($I_p$ is the ionization potential, and $U_p=A_0^2/4$ is the ponderomotive energy). The HHG from bulk crystals has been reported recently [4,5] as the development of long-wavelength lasers. It is quite different from the HHG from gases. The cutoff energy of solid HHG depends on the amplitude of the laser field linearly [4]. It also exhibits multi-plateau structure [6]. However, the mechanisms of HHG from solids are much debated. Inter- and intra-band transitions [4,7,8] are proposed. Three-step model in the coordinate space [9] and vector $k$ space [10] are investigated. Recently, multi-plateau HHG from solids are studied theoretically [7,8,11] and experimentally. Our theoretical model [10] propose that the multi-plateau structure reflects the population of higher conduction bands which are pumped step by step.

In the solid systems, even-order harmonics are measured experimentally. They may come from the interference of HHG trajectories in the different valence bands [12]. It may also come from the Berry phase [13]. In this work, we will present a new mechanism for even-order harmonic generation in symmetric periodic potentials.

Redshifts of HHG spectra were predicted in the HHG from gases and plasmas [14–20]. They originate from delayed emission times of HHG from the excited state, resonate state, or dissociation of molecules. It is a nonadiabatic effect. The feature is that most of the HHG signals are generated from the trailing edge of the laser pulse, where $dI/dt<0$. The driving force from the laser fields is weaker than the previous cycle. The electrons obtain less energy compared to the previous cycle and redshifts of HHG occur. They have been experimentally confirmed [18,21–23]. It was reported that redshift was observed in the optical absorption of crystals in the presence of an intense mid-infrared laser field [24]. In this letter, we find that redshift also exists in emission of HHG from solids. It is universal. The higher conduction bands are excited step by step. As a result, the HHG process in the higher plateau is delayed compared to ultrafast process in the lower conduction bands. As will be shown next, even the system is symmetric in the coordinate space, the electron motion is not symmetric in the $k$ space. Consequently, redshifts of HHG and even-order harmonics are predicted in this work.

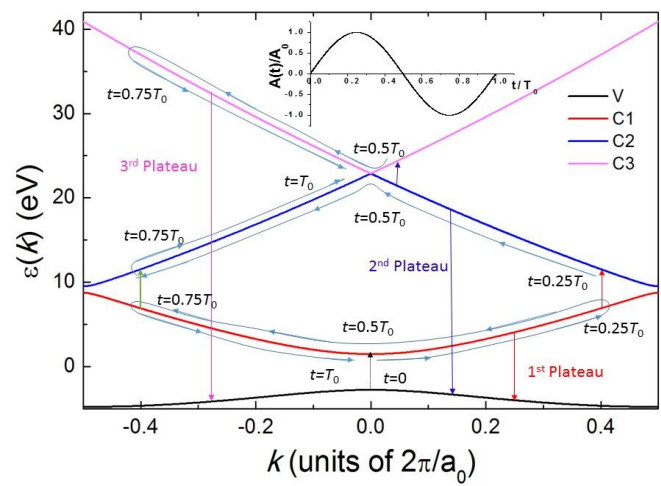

**Fig. 1.** Schematic illustration of the motion of electrons and the multi-plateau HHG in the k space.

The calculations are based on quantum simulations of electron motion in a periodic potential [7,25] $V(x)=-V_0[1+\cos(2\pi x/a_0)]$. $V_0$=0.37 a.u., and the lattice constant $a_0$=8 a.u. The band structure is shown in Fig. 1. For the details of the numerical method we refer readers to Ref. [5]

We briefly review our proposed model of HHG from solids in one laser cycle in the momentum space in strong field approximation [10]. The wave vector

$$k(t) = k_0 + A(t), \quad (1)$$

where $A(t)$ is the vector potential of the laser fields. As illustrated in Fig. 1, at the peak of laser fields ($A(t)$=0), the electron on the valence band V is pumped to the first conduction band C1 with $k\sim0$. After around a quarter cycle $0.25T_0$, part of the electrons are pumped to the conduction band C2 when the bandgap between C1 and C2 is small. The electrons remaining on the valence band C1 oscillate along this band and evolve to the end of cycle and return to its original position $k$=0 at $t=T_0$. This process is symmetric in the $k$ space and the transition from C1 to V is responsible to the first plateau of HHG. Only odd harmonics should be observed. For the electrons pumped to C2 at $t$=0.25$T_0$, they propagate along C2 and may be pumped to C3 at around $t$=0.5$T_0$. The electrons remaining on C2 evolve in the laser fields and return to $k$=0 at the end of this cycle. Part of the electrons on C2 may come from the transition from C1 at around $t$=0.75 $T_0$. This process is asymmetric in the $k$ space, and the transition from C2 to V corresponds to the second plateau of HHG. Even order harmonic should appear due to the break of symmetry. The emission time of HHG from C2 is delayed compared to those from C1. As a result, the spectra of the second plateau in HHG should be redshifted. For the same reason, C3 is populated at around $t$=0.5 $T_0$. It is responsible to the third plateau of HHG. Both the redshifts and even order harmonics should be observed. If we neglect the population of C2 and C3 accumulated in the previous cycles, the number of channels for HHG in the first plateau from the transition of C1 to V is four. The number of channels for the second and third plateaus of HHG is three and two, respectively.

To verify the above analysis, we performed 1-D quantum simulations based on the single-active-electron approximation in the coordinate space. The wavefunctions are expanded by B-spline basis, and the time propagation is based on Crank-Nicholson method. The details can be found in Refs. [8]. The electric field of lasers is expressed as [26]:

$$E(t) = E_0 \cos(\omega(t - \tau/2)) \cos^2(\pi(t - \tau/2)/\tau) \quad (2)$$

where $\tau$ is the total duration of the laser pulse, $E_0$ is the amplitude of field, and $\omega$ is the angular frequency. The wavelength used in this work is 3.2μm, the peak intensity is $I$=8.77x10$^{11}$W/cm$^2$. The HHG spectra with $\tau$ =10 cycles and 15 cycles are presented in Figs. 2 and 3, respectively.

From Figs. 2 and 3 one can find that the HHG spectra exhibit multiple plateaus. This multiple plateau structure agrees with the recent experimental measurement of HHG from solid rare gases [6]. The maximum $k_{max}=A_0$ if we set $k_0=0$ in Eq. (1). The order of the cutoff energy of the first plateau of HHG should be:

$$N_1 = \left[\varepsilon_{C1}(k_{max}) - \varepsilon_v(k_{max})\right]/\omega \approx 32.$$

For the same reason, $N_2$=66, $N_3$=112. One can see that the predicted cutoff energies agree with the TDSE calculations. We also find novel features of HHG in the higher plateaus in this work. From Figs. 2 and 3, only odd-order harmonics are generated in the first plateau. Even-order harmonics appear in the second and third plateaus. This agrees with the expectation of our proposed model in the $k$ space due to the break of symmetry of the electron motion in C2 and C3. However, the intensity of even harmonics is small. The reason is that the population of the conduction bands C2 and C3 by the above step-by-step excitation model is not zero at $t=T_0$. The population will be accumulated to the next cycle. As a result, the strong asymmetric excitation process in the $k$ space approaches to be roughly symmetric in the view of the whole multicycle laser pulse. Weak even-order harmonics also clearly appear in the experimental measurements in the Fig. 1 in Ref. [6], but not discussed. This work sheds light on its mechanism.

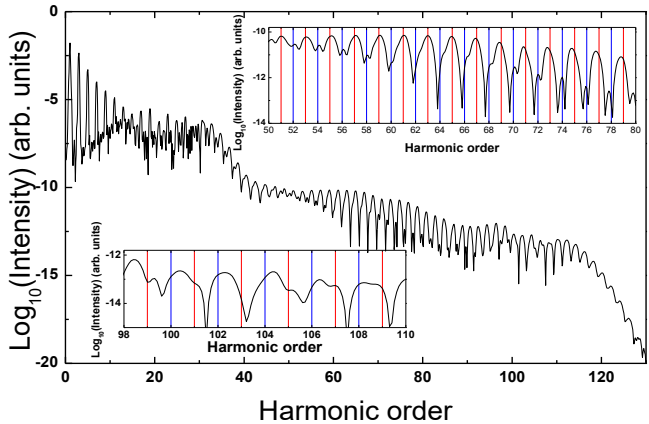

**Fig. 2.** Spectrum of HHG. The parameters of the laser field: $\lambda$ =3.2 μm, I=8.77x10$^{11}$W/cm$^2$, and τ=10 cycles. Part of the spectrum in the second and third plateau is enlarged.

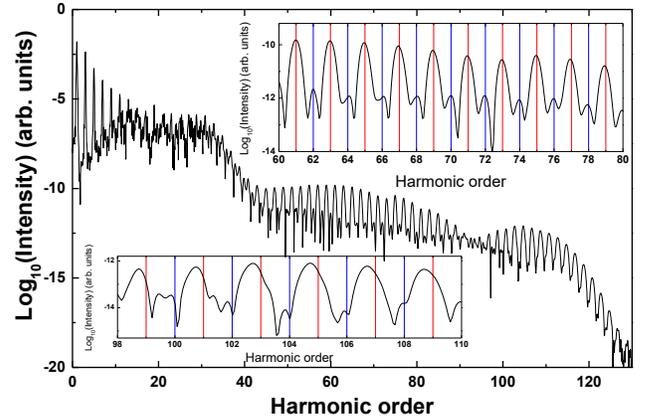

**Fig. 3.** The same as Fig. 3 except τ=15 cycles.

The second feature is that the width of harmonics increases as the order of harmonics increases. It can also be interpreted by the above model. The population of C2 can only be excited when the |E(t)| is big enough. The obvious population of C2 will be delayed for a few optical cycles compared to that of C1 band. For the same reason, the population of C3 is delayed compared to that of C2. As a result, the duration of HHG emission decreases as the harmonic order increases. This can be found by the time-profile analysis of HHG in Ref. [27,28]. From the view of Fourier transformation, shorter duration of signals in the time domain will lead to broader spectrum. This feature uncovers the step-by-step excitation process of HHG in different plateaus.

The third feature is that the spectrum of HHG in Fig. 2 is gradually red-shifted with harmonic order $N>60$ when $\tau=10$ cycles. The shift increases as the order of HHG increases. For HHG in Fig.3 with $\tau=15$ cycles, the spectrum starts to be redshifted in the third plateau. The redshifts come from the nonadiabatic response of electron motion compared to the laser fields. The analysis of the above paragraph can qualitatively explain the redshifts of HHG. The delayed population of C2 and C3 leads to the delay of HHG emission. Most of HHG signals are not symmetric compared to the center of the laser pulse. $dI/dt<0$ in the falling edge of the laser pulse results the redshifts of HHG in the second and third plateau. To quantitatively describe the delayed emission of HHG, we define an asymmetry coefficient [17]:

$$A(N) = \frac{S_f - S_r}{S_f + S_r}, \quad (3)$$

where $S_f$ and $S_r$ are the amount of harmonic $N\omega$ generated in the falling half and rising half parts of the laser pulse, i.e.,

$$S_f = \int_{\tau/2}^{\tau} g(N\omega, t)dt, \quad S_r = \int_0^{\tau/2} g(N\omega, t)dt, \quad (4)$$

$g(N\omega, t)$ is the time profile of harmonics $N\omega$ [27,28]. One see in Fig. 4 that the trend of $A(N)$ agrees well with the amount of redshifts of HHG in Figs. 2 and 3. Longer pulses will reduce the nonadiabatic effects.

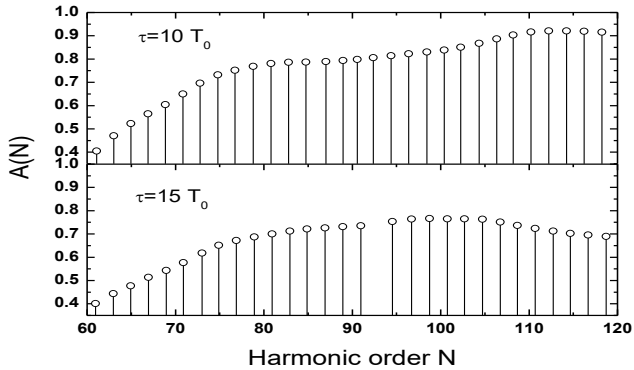

**Fig. 4.** The asymmetric coefficients as a function of harmonic order. In the upper panel, the total pulse duration τ=10 cycles. In the lower panel, τ=15 cycles.

In conclusion, we have studied the spectra of the multi-plateau HHG from solids. Redshifts of HHG in the higher plateau are predicted for the first time. They come from the step-by-step excitation process of higher conduction bands. The width of HHG increases as a function of the harmonic order. We revealed the mechanism of even-order harmonic generation appearing in the higher plateau, which can be used to interpret the recent experimental measurements in Ref. [6]. To our knowledge, this has not been reported. It is hoped that our theoretical work will stimulate further experimental studies on HHG from solids.

**Funding.** National Natural Science Foundation of China (NSFC) (11404376, 11561121002, 11674363, 61377109). Youth Science Foundation of Henan Normal University (2015QK03). Start-up Foundation for Doctors of Henan Normal University (QD15217).
**Acknowledgment**. We thank Tao-Yuan Du and Mu-Zi Li for many very helpful discussions.